\documentclass[10pt,dvips]{article}
\usepackage{amssymb,amsfonts,amsmath,latexsym}
\usepackage{graphics,graphicx,epsf}
\newcommand{\be}{\begin{equation}}
\newcommand{\ee}{\end{equation}}
\newcommand{\bea}{\begin{eqnarray}}
\newcommand{\eea}{\end{eqnarray}}
\newcommand{\ba}{\begin{array}}
\newcommand{\ea}{\end{array}}
\newcommand{\bt}{\begin{tabular}}
\newcommand{\et}{\end{tabular}}

\newcommand{\fr}{\frac}
\newcommand{\ci}{\cite}
\newcommand{\cl}{\centerline}
\newcommand{\bs}{\bigskip}

\newcommand{\vs}{\vspace}

\newcommand{\en}{\eqno}

\newcommand{\fns}{\footnotesize}

\newcommand{\bbib}{\begin{thebibliography}}
\newcommand{\ebib}{\end{thebibliography}}

\newcommand{\und}{\underline}

\newcommand{\Lrar}{\Longrightarrow}
\begin{document}
\bs
\cl{\bf TOPOLOGICAL QUANTIZATION OF CURRENT}
\cl{\bf IN QUANTUM TUNNEL CONTACTS}

\bs

\cl{\bf S.A.Bulgadaev \footnote{e-mail: bulgad@itp.ac.ru}
}
\bs

\cl{\fns Landau Institute for Theoretical Physics,
Chernogolovka, Moscow Region, Russia, 142432}

\bs

\begin{quote}
\footnotesize{It is shown that an account of the Berry phase (a topological $\theta$-term) together with a dissipative term in the effective action $S[\phi]$ of the tunnel contacts induces a strong quantization of the tunnel current at low temperatures. 
This phenomenon like the Coulomb blockade reflects a discrete charge structure of the quantum shot noise and can ensure a quantization
of the tunnel current without a capacitive charging energy $E_C$, when the latter is strongly suppressed by quantum fluctuations.
Since a value of the $\theta$-parameter is determined by the gate voltage,
this effect allows to control a current through the contact.
 A possible physical application of this effect is proposed.}
\end{quote}

\bs
\cl{PACS: 73.23 Hk, 73.40.Gk, 73.21.La, 74.50.+r}
\bs

\underline{\bf 1. Introduction}

\bs

Last decade, in connection with the nanotechnologies, there is a great and permanently growing interest to various small (nanoscale) quantum systems
such as quantum dots, tunnel contacts etc. One of the simplest example
of such systems is a small normal metal tunnel contact - the single electron box (SEB) or the single electron transistor (SET) (see, for example, \ci{1}). As is known, due to the discreteness of the electric  charge of carriers (in this case electrons with a charge e) and a capacitive charging energy $E_c = e^2/2C$,  
the so called Coulomb blockade (CB) takes place in the SEB \ci{2}. It means that the SEB can conduct only when the electrostatic energy $E_C$ of the SEB is equal for states with n and (n+1) electrons. This takes place  when the gate voltage $V_g$ applied to the SEB  is  \ci{2}
$$
V_g = e(n+1/2)/C_g ,
\en(1)
$$
here $C_g$ is a gate capacitance.
This theoretical explanation of the Coulomb blockade is pure classical and is applicable only at moderately low   
temperatures $T < E_C$ and does not take into account quantum phase and/or charge fluctuations. The effective  action of the SEB $S_e[\phi]$ in a phase representation and an imaginary time $\tau$, accounting these fluctuations,  was deduced in \ci{3} (see also \ci{4}) and has the form
$$
S_e[\phi]= S_C[\phi] + S_D[\phi], \quad S_C[\phi]= \int d\tau \fr{{\dot \phi}^2}{4E_C}, 
$$
$$
S_D[\phi]= \iint d\tau d\tau' \alpha(\tau-\tau')\sin^2(\fr{\phi(\tau)-\phi(\tau')}{2}),
\quad 0 \le \tau \le 1/T,
$$
$$
\alpha(\tau) = g \fr{T^2}{\sin^2(\pi \tau T)}, \quad
g = g_t/g_q = R_q/R_t \quad \phi(1/T) = \phi(0) \; (mod \; 2\pi).
\en(2)
$$
Here $g$ is a dimensionless tunnel conductance, $R_q = 1/g_q = h/e^2$ is the quantum resistance and $R_t=1/g_t$ is a tunnel resistance. The action $S_D$ describes a phase correlation and, in some sense, is universal for scale invariant (in $\tau$) systems.  

The action $S_e[\phi]$ is a quite general and with an obvious change of the value of the charge e can be also applied for a description of the phase fluctuations in superconducting tunnel contacts (where it was firstly obtained)\ci{3}, in various
tunnel contacts \ci{4} and granular metals (see, for example, a recent review \ci{5}). It is worth also to note that the CB effect can exist also in the other quantum contacts with different types of scattering \ci{6,7}.

The full partition function, describing the CB effect in the SEB and 
containing also an external gate voltage $V_g,$ has an additional, the Berry phase like, term (the so called $\theta$-term) \ci{4},
$$
Z_{\theta} = \int D\phi \exp(-S[\phi]), \quad
S[\phi] = S_e[\phi] + i S_{\theta}[\phi],
$$
$$
S_{\theta}[\phi]= \theta Q, \quad Q = \fr{1}{2\pi} \int d\tau \dot \phi, \quad
\theta = C_g V_g 2\pi/e. 
\en(3)
$$
where $Q \in \mathbb Z$ is a topological charge of mappings $S^1 \to S^1.$
The $\theta$-term $S_{\theta}[\phi]$ has a pure topological structure.
All physical quantities of the model must be periodic in $\theta$ with a
period $2\pi.$

The SEB model effective action $S_e$  has many properties
similar to those of two-dimensional nonlinear sigma-models (NSM's). It is connected with two facts: 

1) at low T (or at large time scales) the main role plays $S_D$, a dissipative
part of $S_e,$ while the capacitive part $S_C$ is irrelevant \ci{8} in the renorm-group (RG) sense and is strongly suppressed \ci{4}; 

2) on a classical level $S_D$ is invariant under transformations of a small conformal (linear fractional) group \ci{9}. 

Then, it was shown by the renorm-group method that this model is asymptotically free \ci{8,9} and, due to the nontrivial  topology of the unit circle $S_1$, where a phase $\phi$ takes its values, has instanton solutions \ci{9,10}.

The further intensive investigations of the CB properties of the SEB model, using various approximations and methods (from quasi-classical till numerical, see, for example, [11-14]), have given many interesting results (including the renormalized charging energy
$\widetilde E_C$  and a possibility of a phase transition from a conducting to insulating state \ci{4,11}), though some of them differ from each other quantitatively (for $\widetilde E_C$) and even qualitatively (for a phase transition).  Since the CB effect is connected with charge energy
$E_C,$ the larger part of these studies was devoted to its renormalization and all of them confirm a strong suppression of $E_C$ at low T (or large t).
But, a strong suppression of the charging energy $E_C$ induces a question: how a discreteness of the charge can show up itself in this case? 
In particular, can one see the corresponding oscillations of the tunnel current
at $T>\widetilde E_C,$ when they must be washed out by thermal fluctuations?

In this letter,  I  propose a resolution of this dilemma basing
on topological and symmetrical properties of $S.$ It will be shown that
at low $T$ (or large time scales)
a discreteness of the charge and current in the SEB can be realized {\it without the Coulomb blockade}, but due to a presence of the $\theta$-term only. 

\bs

\underline{\bf 2. Topology, symmetries and correlations in $D \le2$}

\bs

A decisive role of  a topology and symmetries  in a determination
of the dynamical and correlation properties of two-dimensional physical systems is known more than 30 years and can be seen in a framework of the effective action, containing in the simplest form
the main ingredients connected with a topology and symmetries.
In the most often cases, when only the lowest fluctuations are important, these  effective actions are the actions of NSM's with the corresponding topological and symmetrical properties.
Now the NSM's are widely used in different physical problems. The most known example is the NSM description of the integer quantum Hall effect (IQHE) in terms of the NSM on the Grassmanian manifold $G=U(2N)/[U(N)\times U(N)]$ (before taking a replica limit $N\to 0$)[15]. It incorporates two important ingredients, the weak localization theory (WLT) and an existence of instanton solutions, and can describe a phase transition. 
It was noted also in this theory that, in general, $\theta$ is not independent  parameter, but can be renormalized by fluctuations (see, for example, \ci{16,17}). Moreover, it was  more or less established for various 2D NSM's with $\theta$-term that an account of $S_{\theta}$ can give the massless (power law) correlations at special values of the $\theta$-parameter \ci{18,19}
$$ 
\theta = \pm 2\pi (n+1/2) \quad (n=0,1,2,...).
\en(4)
$$
It means that at these values of $\theta$ the NSM is in other, massless, phase, while at other values of $\theta$ it has only massive correlations.

This result was firstly conjectured on the physical grounds alone \ci{18}, but later they were confirmed by different methods \ci{19}, including the exact solution ones \ci{20}.  However, one must note that an exact nature of the
appearance of the massless phases at the points  $\theta = \pm 2\pi (n+1/2)$ still remains obscure. As it follows from (3), the partition function  $Z_{\theta}$ has an alternating structure  in this case 
$$
Z_{\theta =\pi} = \sum_{n=-\infty}^{\infty} (-1)^n Z_n,  
\en(5)
$$
where $Z_n$ is the partition function for phase configurations with $Q=n.$
For this reason, the appearance of the massless phases at these points can be connected with a some deep {\it topological} property of the functional space of the NSM's, admitting instantons, and possibly  reflects a larger symmetry of the models at these points, which contains or reduces to a conformal symmetry group as a subgroup.

The action $S$ from (3), though is effectively  one-dimensional, 
also belongs to this kind of actions due to its two properties listed above. 
In our further consideration an account of the $S_{\theta}$  will  play an important role. 
As it follows from (3) all physical quantities
of the SEB model must satisfy the next symmetries:

1) they must be periodic in $\theta$ with a period equal to $2\pi$ \; 
(thus, one can confine oneself by a range $0 \le \theta \le 2\pi$);

2) they  divide into even and odd parts relative $\theta$.

Recently, it was shown under some assumptions that $S_e$ is exactly solvable
and a series of higher order conserved currents was constructed \ci{21}. Moreover, it was shown also that the point $\theta = \pi$ has some special properties, corresponding to a massless phase.  Thus, one can see that our one-dimensional
model also has a massless fixed point at $\theta = \pi.$
This allows us to analyze our model (3) basing on its topological and
above listed properties.

\bs

\underline{\bf 3. RG phase diagram and charge quantization}

\bs

Now I will consider the RG flow phase diagram in the $g,\theta$-plane using above mentioned facts.
In particular, a renormalized physical conductance $g$ and its $\beta$-function
$\beta_g(g,\theta)$, determining its change under RG flow, are an even function of $\theta,$ while the corresponding $\beta$-function of $\theta,$ $\beta_{\theta}(g,\theta),$ must be odd in $\theta.$ Since they must also be periodic in $\theta,$ one can write the following functional representation for them 
$$
d_l g = \beta_g(g,\theta), \quad
\beta_g(g,\theta)= a_0(g) + \sum_1^{\infty} a_n(g) \cos(n \theta),
\en(6)
$$
$$
d_l \theta = \beta_{\theta}(g,\theta), \quad
\beta_{\theta}(g,\theta)= \sum_1^{\infty} b_n(g) \sin(n \theta), 
\quad l = \ln \tau/\tau_0.
\en(6')
$$
Here $\tau = \tau_0 e^l$  is a running scale with $\tau_0$ fixing an initial scale, which is in our case a scale of the SEB. The periodic parts appear due to the large phase fluctuations, the instantons (or phase slips) contributions, in an analogy with the two-dimensional IQHE case \ci{17}.
The function $a_0(g)$ is now known up to two-loop level \ci{14}
$$
a_0(g)= -(1+1/g).
\en(7)
$$
One can write, for instance, the one-instanton contribution. It has a form 
$$
a_1(g) = - D g^2 \exp(-g) = b_1(g),
\en(8)
$$
where the constant $D > 0$ is connected with the fluctuations
over one instanton solution. Note that one instanton contribution in $\beta_g$ enhances a decreasing of g at $\theta = 0$ and slows down it at $\theta = \pi.$
The renormalization of $\theta$ is absent in the RG approach at small 1/g and contains only the instanton contributions. This coincides with the IQHE case,
where $a_0$ has a similar form before taking a replica limit $N \to 0$. 

As follows from (6') the lines 
$\theta = \pm n \pi \; (n=0,1,2,...) $ are integral lines, since
they correspond to zeros of $\beta_{\theta}$. On the line $\theta =0$ (and on 
other $\theta = 2n\pi$ lines) one has usual asymptotic freedom behaviour with a trivial infra-red (IR) stable fixed point $g=0.$  In the limit 
$\tau \to \infty$ or $T \to 0$ the SEB does not conduct  on this line at all.

On the line $\theta = \pi$ there must be the IR-stable
finite fixed point $g^*$, corresponding to the massless phase, which ensures a charge transfer in the SEB at $\theta = \pi$. It means that the SEB
conducts on this line at arbitrary initial values $g_0.$ This means that on this line at low T  the initial tunnel conductance $g_0$ is firstly reduced logarithmically by the small quantum phase fluctuations (like in the WLT) till $\tau_c \sim \tau_0 \exp(g),$ a correlation imaginary time (or a characteristic temperature $T_c \sim 1/\tau_c$). The correlation time $\tau_c$ is similar simultaneously to the dephasing time $\tau_{\phi}$ and a correlation length $L_c$ of the WLT, since in the model (3) there exists only one effective dimension, $\tau.$ Further, at $\tau > \tau_c,$ a renormalized conductance $g \to 0$ exponentially  $g \sim e^{-\tau/\tau_c}$  due to the large phase fluctuations - instantons (or phase slips). Analogously,  $g \sim e^{-T_c/T}$ at $T_c > T.$
In the limit 
$\tau \to \infty$ or $T \to 0$ the SEB does not conduct on this line at all.

On the line $\theta = \pi$ a behaviour of $g$ differs from that for $\theta = 0.$ There must be the 
finite fixed point (FP) $g^*$, corresponding to the massless phase, which is  IR-stable for initial values of $g$ on this line and IR-unstable for initial values of $g$ out of this line. This FP ensures a charge transfer in the SEB at $\theta = \pi$. Such behaviour of $g$ means that the SEB
conducts on this line at arbitrary initial values $g_0.$

Unfortunately, the exact value of $g^*$ is now not known. Here is some difficulty. $a_n(g) \; (n > 0)$, entering in $\beta_g,$ contain the constants like $D$, which are not defined unambiguously: they depend on scheme of renormalization \ci{18}. At the same time the condition of the existence of only one nontrivial fixed point is very restrictive. To see this let us consider
one instanton contribution. Then one obtains the equation
$$
\beta_g(g^*) = a_0 - a_1 = 0 \quad \Lrar \quad (1+\fr{1}{g^*}) = D {g^*}^s \exp(-g^*).
\en(9)
$$
Here, for a generality, we consider a possibility, when the power of g in $a_1$
is some integer $s= 1,2,3,...$.  
The left side of (9) is a smooth function convex down, while the right side is convex up with a maximum at $g=s.$ Then,
neglecting
a smooth dependence of the left side of (9) (its derivative $\sim 1/g^2$), we see that one solution exists only for a
special value of $D$:
$$
g^* = s, \quad
D_{s} = \fr{g^*+1}{(g^*)^{s+1}} e^{g^*}. 
\en(10)
$$
It is very intriguing that in this approximation $g^*$ is an integer.
If the exact value of $g^*=s,$ then it means that  $g_t^* = s g_q = s h/e^2$.
An account of the left side derivative gives 
$$
g^*=\frac{s-1+\sqrt{(s+1)^2+4}}{2}.
\eqno(10')
$$
For larger values of $D$ one obtains two solutions, while for smaller values
there is no any solution. 
A case of two solutions, taken formally, has some difficulties, because two nearest fixed points cannot be both IR-stable on line. Then there are two possibilities: 1) the smaller FP $g^*_1$ is IR-stable and the larger FP $g^*_2$
is IR-unstable or 2) an opposite case.  In the first case  the renormalized $g$ for initial values $0 < g_0 <g^*_2$ tends at $\tau \to \infty $ (or $T\to 0$) to
the FP $g^*_1,$ which will define the effective conductance of the SEB for these
initial values. For $g^*_1 < g_0 < \infty$ it tends to $\infty,$ what would mean that for these initial values the effective conductance of the SEB is $\infty.$
In the second case, when $g^*_1$ is IR-unstable one obtains the effective
conductance $g=g^*_2$ for all $g_0 > g^*_1 $ and $g=0$ for  $g_0 < g^*_1.$ 
Both these results are unacceptable since they contradict to the exact result \ci{21} about one massless phase at $\theta = \pi$. In order the two FP on a line be IR-stable 
one needs one IR-unstable fixed point between them, what is impossible in our case in the region near a maximum of $a_1$. At the same time an existence of two FP could be interpreted as a mathematical signal that physically in the model
there is a  narrow band of the massless states with a width equal to a distance between these solutions (this hypothesis has been proposed in 80ies during an investigation of the IQHE).

An account of higher order terms does not improve a situation (in 2D NSM's as well as in our 1D model), since a presence of the alternating factor $(-1)^n$ in the corresponding equation
$$
\beta_g(g^*) = a_0 + \sum_1^{\infty} (-1)^n a_n =0.
\en(11)
$$
only complicates a situation, because they can partially cancel each other.
For this reason, in the RG approach one needs to know all terms to make unambiguous conclusion about finite nontrivial fixed point. This can be done only
with an usage of some exact results, as it takes place in \ci{6} and in  the exact solution approach of \ci{21}.
\begin{figure}[t]
\begin{picture}(250,120)
\put(50,0){%
\begin{picture}(80,80)
\put(10,0){\vector(0,1){110}}
\put(10,10){\vector(1,0){70}}
\put(15,105){$\theta/2\pi$}
\put(85,10){$g$}
\qbezier(10,10)(40,25)(10,40)
\qbezier(10,40)(40,55)(10,70)
\qbezier(10,70)(40,85)(10,100)
\put(-10,10){0}
\put(-10,40){$1$}
\put(-10,70){$2$}
\put(25,25){\circle*{3}}
\put(25,55){\circle*{3}}
\put(25,85){\circle*{3}}
\put(70,10){\vector(-1,0){30}}
\put(40,10){\line(-1,0){30}}
\put(70,40){\vector(-1,0){30}}
\put(40,40){\line(-1,0){30}}
\put(70,70){\vector(-1,0){30}}
\put(40,70){\line(-1,0){30}}
\put(70,25){\vector(-1,0){30}}
\put(40,25){\line(-1,0){15}}
\put(70,55){\vector(-1,0){30}}
\put(40,55){\line(-1,0){15}}
\put(70,85){\vector(-1,0){30}}
\put(40,85){\line(-1,0){15}}

\put(17,25){\line(1,0){7}}
\put(10,25){\vector(1,0){8}}
\put(17,55){\line(1,0){7}}
\put(10,55){\vector(1,0){8}}
\put(17,85){\line(1,0){7}}
\put(10,85){\vector(1,0){8}}
\put(32,-18){({\small a})}
\end{picture}}

\put(210,0){%
\begin{picture}(80,80)
\put(10,0){\vector(0,1){100}}
\put(10,10){\vector(1,0){70}}
\put(15,100){$\theta/2\pi$}
\put(90,10){$g$}
\qbezier(10,10)(50,40)(10,70)
\put(-10,10){0}
\put(-10,40){$\fr{1}{2}$}
\put(-10,70){$1$}
\put(30,40){\circle*{3}}
\put(70,10){\vector(-1,0){30}}
\put(40,40){\line(-1,0){10}}
\put(70,40){\vector(-1,0){30}}
\put(70,70){\vector(-1,0){30}}
\put(40,70){\line(-1,0){30}}
\put(70,30){\vector(-1,0){20}}
\put(70,50){\vector(-1,0){20}}
\qbezier(10,10)(23,17)(27,22)
\qbezier(27,22)(35,30)(50,30)
\qbezier(10,70)(23,63)(27,58)
\qbezier(27,58)(35,50)(50,50)
\put(10,40){\vector(1,0){10}}
\put(20,40){\line(1,0){10}}
\put(32,-18){({\small b})}
\end{picture}}
\end{picture}

\vs{1cm}

{\small Fig.1. (a) A schematic picture of the SEB $g-\theta$-plane 
phase diagram at T=0. The arrows on the horizontal lines $\theta/2\pi = 0, \pm 1/2, \pm 1, \pm 3/2,...$ indicate the RG flow.
(b) The same zoomed picture for one period part. Here the RG flow shows that $\theta$
is also renormalized at small $g$ and $\theta/2\pi \ne \pm (n+1/2) \; (n=0,1,2,...).$
}
\end{figure}
Nevertheless, basing on the abovementioned exact properties, one can conjecture a following global RG flow phase diagram of the SEB model at $T \to 0$ (fig.1). Here it is supposed that there are no any other fixed points.
The RG flow near the fixed points and separatrices, connecting these fixed points,
shows a strong renormalization of the parameter $\theta.$ 
If the parameter $\theta$ were unrenormalized, all RG flow lines would be straight lines. 
Then a separatrice, connecting the trivial fixed point $g=0= \theta$ with $g=g^*,
\theta = \pi$  would 
be a continuous line of the fixed points. This would mean that the SEB model is conducting at all $\theta$ except $\theta=0$ and critical correlations have power law form with the exponents depending on $\theta.$ This is not a case in the 2D NSM's and contradicts to the results \ci{21}.

The presented RG flow diagram is very similar to that in the theory of the IQHE \ci{17,18}, since they both are a consequence of general symmetries, topology and an existence of one finite fixed point only.   
The renormalized $\tilde \theta$ has the jumps at $\theta = \pi,$ like the sign
function (fig.2a), while a behavior of the conductance $g$ is similar to a behavior of $\sigma_{xx}.$  
\begin{figure}[t]
\begin{picture}(250,120)
\put(20,0){%
\begin{picture}(100,100)
\put(10,10){\vector(0,1){100}}
\put(10,10){\vector(1,0){110}}
\put(120,10){$\theta/2\pi$}
\put(20,100){$\tilde \theta /2\pi $}
\put(8,-5){0}
\put(25,-5){$\fr{1}{2}$}
\put(47,-5){$1$}
\put(65,-5){$\fr{3}{2}$}
\put(87,-5){$2$}
\put(105,-5){$\fr{5}{2}$}
\put(30,10){\line(0,1){2}}
\put(50,10){\line(0,1){2}}
\put(70,10){\line(0,1){2}}
\put(90,10){\line(0,1){2}}
\put(110,10){\line(0,1){2}}

\put(10,30){\line(1,0){2}}
\put(10,50){\line(1,0){2}}
\put(10,70){\line(1,0){2}}
\put(10,90){\line(1,0){2}}

\put(-10,30){$\fr{1}{2}$}
\put(-10,50){$1$}
\put(-10,70){$\fr{3}{2}$}
\put(-10,90){$2$}
\put(30,30){\circle*{2}}
\put(30,50){\line(1,0){40}}
\put(70,70){\circle*{2}}
\put(70,90){\line(1,0){40}}
\put(65,-25){({\small a})}
\end{picture}
}
\put(190,0){%
\begin{picture}(110,100)
\put(10,10){\vector(0,1){100}}
\put(10,10){\vector(1,0){110}}
\put(120,10){$\theta/2\pi$}
\put(20,100){$\bar n (\theta)$}
\put(8,-5){0}
\put(25,-5){$\fr{1}{2}$}
\put(47,-5){$1$}
\put(65,-5){$\fr{3}{2}$}
\put(87,-5){$2$}
\put(105,-5){$\fr{5}{2}$}
\put(30,10){\line(0,1){2}}
\put(50,10){\line(0,1){2}}
\put(70,10){\line(0,1){2}}
\put(90,10){\line(0,1){2}}
\put(110,10){\line(0,1){2}}

\put(10,30){\line(1,0){2}}
\put(10,50){\line(1,0){2}}
\put(10,70){\line(1,0){2}}
\put(10,90){\line(1,0){2}}

\put(-10,30){$\fr{1}{2}$}
\put(-10,50){$1$}
\put(-10,70){$\fr{3}{2}$}
\put(-10,90){$2$}
\put(30,10){\line(0,1){40}}
\put(30,50){\line(1,0){40}}
\put(70,50){\line(0,1){40}}
\put(70,90){\line(1,0){40}}
\put(65,-25){({\small b})}
\end{picture}}
\end{picture}

\vs{1cm}
\cl{{\small Fig.2.  A schematic picture of (a) $\tilde \theta(\theta)$  and (b) $\bar n(\theta)$ at $T \to 0.$}}
\end{figure}
The average number of electrons in the SEB grain $\bar n$ has a sharp, stepwise, dependence on $\theta$ like the Hall conductivity $\sigma_{xy}$
on magnetic field (fig.2b). 
Using the results of \ci{15}, I have supposed also that  the RG flow  has a linear form near trivial fixed points $g=0, \; \theta = 2n\pi.$ When a perturbation with a some characteristic scale $\tau_p$ is introduced or 
T is small, but finite (or t is large, but finite), the renormalization must
be stopped at scale $\max[T,1/\tau_p]$ on an energy scale or at $\min[1/T,\tau_p]$ on a $\tau$ scale. This gives a smeared behaviour near the lines $\theta = 2\pi(n+1/2).$ For instance, the steps
on fig.2 will be more rounded and smooth. 

Finally, one can see that the obtained topological  charge quantization does not depend on $E_C$ and works even if the renormalized $\widetilde E_C < T,$ when the CB is already impossible: it will be washed out by the thermal fluctuations.  For this reason one obtains at $\widetilde E_C < T$ a quantized (or oscillating) behaviour, smeared by finite $T,$ which takes place completely due to the topological quantization. Since $\widetilde E_C \ll E_C,$ there is a wide temperature
interval $\widetilde E_C \ll T \ll E_C,$ where the charge quantization can 
demonstrate itself through a topological quantization. The existence of this
temperature interval allows to verify the proposed theory of the charge quantization and the corresponding oscillations. 
At the same time the CB can still show up itself, since one can treat 
$\widetilde E_C$ as one of $1/\tau_p.$ Then at low T one must stop renormalization in an energy scale at 
max$[T, \widetilde E_C],$ and for $\widetilde E_C \ge T$ (if this region exists) one can see the smeared CB oscillations.
\bs

\underline{\bf 4. Conclusions and discussions}

\bs
Thus, using topological and symmetrical
properties of the SEB model, it is shown that the quantization of charge
and current in tunnel contacts is possible  even when
the Coulomb charging energy $E_C$ is strongly suppressed by quantum fluctuations
to $\widetilde E_C \le T$. It means that the charge quantization and the corresponding oscillation behaviour of various characteristics of the SEB model 
will survive at low $T \ll E_C$, but a smearing of these functions will be determined by $\max[T, \widetilde E_C].$
 
It is interesting that a qualitatively similar phase diagram picture was obtained for the quantum point contact connecting quantum dot with external lead \ci{6}.
In this case the corresponding model Hamiltonian describes from very beginning
the fermions and  the stable fixed points at $\theta/2\pi = n +1/2$ are identified with the Toulouse fixed point in the theory of the 2-channel Kondo model. One can hope that the obtained results can be expand also on quantum contacts with other types of scattering \ci{7}.

Though the $\theta$-parameter depends on the gate voltage $V_g,$ it turns out
that really the physics of the SEB model at low T weakly depends on its value, since  effectively $\theta$ is strongly renormalized to $0 \, (mod \; 2\pi)$ for all $\theta$ except $\theta = \pm 2\pi (n+1/2).$

At the end of this letter a few words about a possible physical application
of the obtained results.
An useful quantum device connected with the SEB model is the SET. 
Then the obtained topological quantization of the charge in the SEB model 
at $\widetilde E_C \le T$ can 
be observed as a topological quantization of the tunnel current in the SET.
 
\bs

\und{ Acknowledgments}

\bs
The author is thankful to his friends for useful and fruitful discussions and 
conversations.
The work was partly supported by the RFBR grants 02-02-16403, 2044-2003-2 and by the ESF network AQDJJ.


\bbib{50}
\bibitem{1} "Single Charge Tunneling", ed.by H.Grabert and M.H.Devoret, NATO ASI, Ser.B, v.294, (Plenum, New York, 1992).
\bibitem{2} D.V.Averin, K.K.Likharev, in "Mesoscopic Phenomena in Solids", Amsterdam, Elsevier, (1991) 173.
\bibitem{3} V.Ambegaokar, U.Eckern and G.Sch\"on, Phys.Rev.Lett.{\bf 48} (1982) 1745. 

\bibitem{4} G.Sch\"on, A.D.Zaikin, Physics Reports {\bf 198} (1990) 237.
\bibitem{5} I.S.Beloborodov, A.V.Lopatin, V.M.Vinokur, K.B.Efetov, cond-mat/0603522.
\bibitem{6} K.A.Matveev, ZhETF {\bf 99} (1991) 1598; Phys.Rev. {\bf B51} (1995) 1743.
\bibitem{7} Yu.V.Nazarov, Phys.Rev.Lett. {\bf 82} (1999) 1245.

\bibitem{8} S.A.Bulgadaev,  Pis'ma v ZhETF {\bf 45} (1987) 486 (Sov.Phys. JETP 
{\bf 32} (1987) 348).
\bibitem{9} S.A.Bulgadaev, Phys.Lett. {\bf A125} (1987) 299.
\bibitem{10} S.V.Korshunov, Pis'ma v ZhETF {\bf 45} (1987) 449.

\bibitem{11} S.V.Panyukov, A.D.Zaikin, Phys.Rev.Lett. {\bf 67} (1991) 3168.
\bibitem{12} G.Falci, G.Sch\"on, and G.T.Zimanyi, Phys.Rev.Lett. {\bf 74} (1995) 3257;
\bibitem{13} X.Wang, H.Grabert, Phys.Rev. {\bf B53} (1996) 12621.
\bibitem{14}  W.Hofstetter, W.Zwerger, Phys.Rev.Lett. {\bf 78} (1997) 3737.

\bibitem{15} H.Levine, S.B.Libby and A.M.M.Pruisken, Phys.Rev.Lett. {\bf 51} (1983) 1915.
\bibitem{16} D.E.Khmel'nitskii,  Pis'ma v ZhETF {\bf 45} (1983) 486.
\bibitem{17} A.M.M.Pruisken, Phys.Rev. {\bf B32} (1985) 2636.

\bibitem{18} F.D.M.Haldane, Phys.Rev.Lett. {\bf 50} (1983) 1153.
\bibitem{19} I.Affleck, F.D.M.Haldane, Phys.Rev. {\bf B36} (1987) 5291.
\bibitem{20} A.B.Zamolodchikov, Al.B.Zamolodchikov, Nucl.Phys. {\bf B379} (1992) 602.
\bibitem{21} S.L.Lukyanov, A.B.Zamolodchikov, J.Stat.Mech.:Theor.Exp. P05003
 (2004); hep-th/0306188.

\ebib
\end{document}